\documentclass[twocolumn]{aastex701}

\usepackage{amsmath}
\usepackage{graphicx}
\usepackage{bm}
\usepackage{hyphenat}
\usepackage{soul}

\begin{document}

\title{Non-Markovian Cosmic-Ray Pitch-Angle Transport from Mirror Interactions}

\author{Kai Yan}
\email{kaiyan@smail.nju.edu.cn}
\affiliation{School of Astronomy and Space Science, Nanjing University, Nanjing 210023, China}
\affiliation{Key laboratory of Modern Astronomy and Astrophysics (Nanjing University), Ministry of Education, Nanjing 210023, China}
\affiliation{Institut für Physik und Astronomie, Universität Potsdam, D-14476 Potsdam, Germany}

\author{Huirong Yan}
\email[show]{huirong.yan@desy.de}
\affiliation{Deutsches Elektronen\hyp{}Synchrotron (DESY), Platanenallee~6, D-15738 Zeuthen, Germany}
\affiliation{Institut für Physik und Astronomie, Universität Potsdam, D-14476 Potsdam, Germany}

\author{Parth Pavaskar}
\email{parth.pavaskar@desy.de}
\affiliation{Deutsches Elektronen\hyp{}Synchrotron (DESY), Platanenallee~6, D-15738 Zeuthen, Germany}
\affiliation{Institut für Physik und Astronomie, Universität Potsdam, D-14476 Potsdam, Germany}

\author{Chuanpeng Hou}
\email{chuanpeng.hou@uni-potsdam.de}
\affiliation{Institut für Physik und Astronomie, Universität Potsdam, D-14476 Potsdam, Germany}

\author{Ruo-Yu Liu}
\email[show]{ryliu@nju.edu.cn}
\affiliation{School of Astronomy and Space Science, Nanjing University, Nanjing 210023, China}
\affiliation{Key laboratory of Modern Astronomy and Astrophysics (Nanjing University), Ministry of Education, Nanjing 210023, China}


\begin{abstract}
Cosmic-ray pitch-angle transport in magnetohydrodynamic (MHD) turbulence is governed by the interplay between magnetic mirroring and gyroresonant scattering. We develop a guiding-center (GC) Langevin model with explicit mirror drift and gyroresonant diffusion to describe the pitch angle evolution. This model accurately captures our test-particle simulation results in three-dimensional MHD turbulence, driven both solenoidally and compressively. We find that magnetic mirroring can drive anomalous pitch-angle diffusion at large pitch angles (including $90^\circ$) with non-Markovian memory effects, which arises from trapping of particles in magnetic wells. 
Gyroresonant scattering controls the escape rate from these wells. Across $M_{\rm A}$, large-pitch-angle particles are jointly regulated by mirror trapping and gyroresonant escape, exhibiting a transition from anomalous to normal diffusive pitch-angle transport as scattering strengthens, whereas small-pitch-angle particles remain gyroresonance-dominated and diffusive throughout. The pitch angle transport is found to be dominated by the compressible perturbations with marginal influence from Alfv\'en modes. In compressible turbulence with realistic damping accounted for, transit time damping (TTD) treatment fully recovers mirror interactions.
\end{abstract}

\section{Introduction}
Cosmic-ray (CR) transport in magnetized turbulence underpins particle acceleration and nonthermal emission across astrophysical and space-plasma environments. In practice, CR propagation is often described as a diffusive process: it is assumed that pitch-angle scattering rapidly decorrelates particle directions, allowing transport to be modeled by a Markovian diffusion equation with well-defined coefficients\citep{1966Jokipii, 2002Schlickeiser}. However, when scattering is intermittent or insufficient, particle trajectories retain memory of the underlying magnetic structure, rendering a diffusive description invalid\citep{2020Zimbardo, 2022Maiti, zhao2025a, 2025Lubke}. 

A key development beyond standard quasilinear theory (QLT) was the recognition that near-$90^{\circ}$ dynamics is strongly affected by compressive fluctuations and non-linear mirror effects\citep{2001Felice, 2008Yan}. The $\delta$-function resonance function of QLT yields vanishing pitch-angle scattering near $90^{\circ}$, and this limitation has long been understood as an artifact of the unperturbed guiding-center orbit assumption and the resulted idealized resonance function\citep{1966Jokipii,1969Kulsrud,2014Qin}. In particular, compressive fluctuations enable  can induce bounce motion and transient trapping through conservation of the first adiabatic invariant $\mu_m=m v_\perp^2/(2B)$  or the resonant mirror interaction through the Landau resonance, the transit-time damping (TTD) process \citep{1973Cesarsky,2001Felice, 2002Schlickeiser}. The conservation of magnetic moment naturally broaden the idealized QLT condition by a finite-width kernel, making TTD effective near $\mu\simeq0$ and alleviating the $90^\circ$ barrier \citep{2008Yan}. Despite broad consensus on the importance of compressive structures, most existing approaches remain within a diffusive framework, in which mirror effects and compressions are typically incorporated as modified scattering rates within a Markovian transport picture. Recent studies has further renewed interest in the interplay between magnetic mirroring and pitch-angle scattering\citep{2025LopezBarquero, 2025Bell, 2025Lubke}.

In this work we demonstrate that mirror structures can drive a distinct regime, namely, mirror-induced anomalous diffusion that lies beyond both QLT and its nonlinear extensions. Rather than postulating a phenomenological memory kernel, we identify this behavior directly from first-principle test-particle dynamics in three-dimensional MHD turbulence. We introduce a guiding-center (GC) Langevin model in which mirror interactions enter as a deterministic drift in $\mu$, while gyroresonance is represented as a stochastic, Markovian diffusion term. This decomposition is intended to describe the regime in which mirror trapping and gyroresonance remain scale-separated.
We show that this model accurately reproduces the test-particle results. At the ensemble level the Langevin model corresponds to a focused Fokker-Planck equation with a spatially varying focusing term, extending constant-focusing treatments \citep{1976Earl, 2014Tautz,2017Lasuik} to the multiscale, intermittent focusing, which is naturally produced by compressive turbulence. The combination of mirror effects and gyroresonant scattering has two linked consequences that regulates the global transport. First, it enables particles to traverse $\mu\simeq 0$ ($90^{\circ}$ barrier) via a deterministic drift. Second, because this drift within a local well is reversible and weakly accumulative, it can also confine particles in intermittent mirror wells, yielding heavy-tailed trapping times and thus non-Markovian, anomalous pitch-angle evolution. In this framework, particle transport is controlled by the competition between mirror trapping and gyroresonant decorrelation. By directly linking anomalous transport to a concrete microscopic mechanism, we aim to identify the conditions (e.g., Alfv{\'e}n Mach number $M_{\rm A}$) under which diffusive description are acceptable and explicit non-Markovian effects must be considered. It offers microscopic insights into CR transport across scales, from early-time propagation after escaping CR accelerators \citep{1974Earl, Prosekin2015,2021Recchia} to the formation of spatially extended nonthermal emission around accelerators \citep{abeysekara2017extended, aharonian2021extended, 2024LHAASOCygnus, 2019Liu}. 

\section{Method}
\subsection{MHD turbulence}
We generate 3D isothermal ideal-MHD turbulence using the \textsc{Athena++} code within a triply periodic box of size $L$ at a resolution of $N^3 = 1024^3$ \citep{2020Stone}. Energy is injected at the scale $k_{\rm inj} = 2$. Turbulence is continuously driven by an Ornstein–Uhlenbeck process $\mathbf{F}^{\rm turb}$, where the solenoidal/compressive mixture is controlled via the projection operator $P^{\zeta}_{ij} = \zeta P^\perp_{ij} + (1-\zeta)P^\parallel_{ij}$ \citep{MakwanaYan2020, 2025Kempski}. We set $\zeta = 0.1$ to for compressive driving datacubes, and $\zeta = 1$ for solenoidal driving datacubes. We adopt a plasma beta of $\beta = 0.5$. The Alfvén Mach number $M_{\rm A}$ is varied by adjusting the energy injection rate. The corresponding sonic Mach number is given by $M_{\rm s} = M_{\rm A}\sqrt{2/(\hat{\gamma}\beta)} = 2M_{\rm A}$ for isothermal plasma ($\hat{\gamma}=1$). In each run, data are collected only after the turbulence reaches a statistically steady state.

\subsection{Test-particle integration}
We trace CR trajectories by integrating the motion of relativistic test particles in the 3D MHD turbulence using the Boris push method with periodic boundaries \citep{2007Press}. Since the speed of a relativistic particle greatly exceeds the Alfv\'en speed, the magnetic field is treated as stationary and the induced electric field is neglected\citep{BYL11, 2013Xu, 2016Cohet, 2022Maiti}. The particle motion follows the Lorentz force equation. The gyrofrequency and Larmor radius are $\Omega_0=qB_0/(\gamma mc)$ and $r_{\rm L}=\gamma mvc/(qB_0)$, respectively. $q$ is particle charge. $m$ is rest mass. $c$ is the speed of light. $v$ is the particle speed. $\gamma$ is the Lorentz factor. $B_0$ is the mean magnetic field. We set $r_{\rm L}=5 \, l_{\rm grid}$ in default, where $l_{\rm grid}$ is an MHD-normalized unit of the grid spacing. This choice places the corresponding resonant wavenumber to be $k = N/5/(2\pi)\approx 32$, which is well within the inertial range of the turbulence (Appendix Fig.\ref{fig:PSD}).

\subsection{Guiding-center Langevin model}
To disentangle mirror effects from gyroresonant scattering, we construct a guiding‑center (GC) Langevin model that describes the evolution of the pitch‑angle cosine $\mu$ as a drift–diffusion process \citep{1999Zhang}, i.e.,
\begin{equation}
d\mu=
-\frac{1-\mu^2}{2B}\,v\,\nabla_s B\,dt
+\frac{\partial D^G_{\mu\mu}}{\partial\mu}\,dt
+\sqrt{2D^G_{\mu\mu}(\mu)}\,\eta_t,
\label{eq:LGV_mu_sde}
\end{equation}
where $\nabla_s = \hat{\mathbf{b}}\cdot\nabla$ (with $\hat{\mathbf{b}}=\mathbf{B}/B$) denotes the gradient along the local magnetic field, and $\eta_t$ is a Markovian and Gaussian white noise satisfying $\langle \eta(dt_1) \eta^*(dt_2) \rangle = \delta(dt_1 - dt_2) \langle \eta(dt)\eta^*(dt) \rangle$, and $\langle \eta(dt) \rangle = 0, \langle \eta(dt)\eta^*(dt) \rangle = dt$\citep{2025Lubke, 2020Zhang}. The first term represents the deterministic mirror (focusing) drift in the adiabatic regime, where the magnetic-field variation scale satisfies \(\ell \gtrsim r_g\). Micromirror structures with \(\ell \ll r_g\), such as those produced by kinetic instabilities in high-\(\beta\) plasma and possibly relevant in ICM-like environments \citep[e.g.,][]{2025NAReichherzer}, are not included in either our simulations or our theoretical analysis. The last two terms together represent stochastic gyroresonant scattering, characterized by the diffusion coefficient $D^G_{\mu\mu}$, which is computed following Eq.(5) of \citet{2008Yan} by integrating the turbulence spectrum over nonlinearly broadened cyclotron-resonant scales

\begin{equation}
\begin{split}
D_{\mu\mu}^G
&= \frac{\Omega_0^2(1-\mu^2)}{B_0^2}
\int d^3k \sum_{n \ge 1}
R_n(k_\parallel v_\parallel-\omega\pm n\Omega_0) \\
&\quad \times
\left[
I^A(\mathbf{k})\frac{n^2J_n^2(W)}{W^2}
+ \frac{k_\parallel^2}{k^2}{J_n'}^2(W)I^M(\mathbf{k})
\right],
\end{split}
\label{eq:LGV_mu}
\end{equation}

where $v_{\parallel}$ ($v_{\perp}$) is the particle speed parallel (perpendicular) to $B_0$, $k_{\parallel}$ ($k_{\perp}$) is the parallel (perpendicular) wavevector, $J_n$ ($J_n'$) is the $n$th-order Bessel function (and its derivative), $R_n$ is the $n$th-order resonant function, $W = k_\perp v_\perp/\Omega_0$, and $I^{A}$ and $I^{M}$ are the scalar energy spectra for Alfv\'enic and magnetosonic modes, respectively. 

The GC position $\mathbf{R}$ evolves as \citep{2006Howes}
\begin{equation}
d\mathbf{R}=v_\parallel \frac{\mathbf{B}}{B}\,dt -\frac{c\,\mu_m}{q}\,\frac{\nabla B\times \mathbf{B}}{B^2} \,dt,
\label{eq:LGV_R}
\end{equation}
where $\mu_m = m v_{\perp}^2/(2B)$ is the magnetic moment. $v_{\parallel}$ ($v_{\perp}$) is the particle speed parallel (perpendicular) to $\mathbf{B}$. By integrating the coupled Langevin equations (\ref{eq:LGV_mu_sde}) and (\ref{eq:LGV_R}), we advance the GC trajectories in the 3D MHD turbulence fields.

\section{Results}
\label{sec:results}

\begin{figure*}[htbp]
\includegraphics[width=0.7\textwidth]{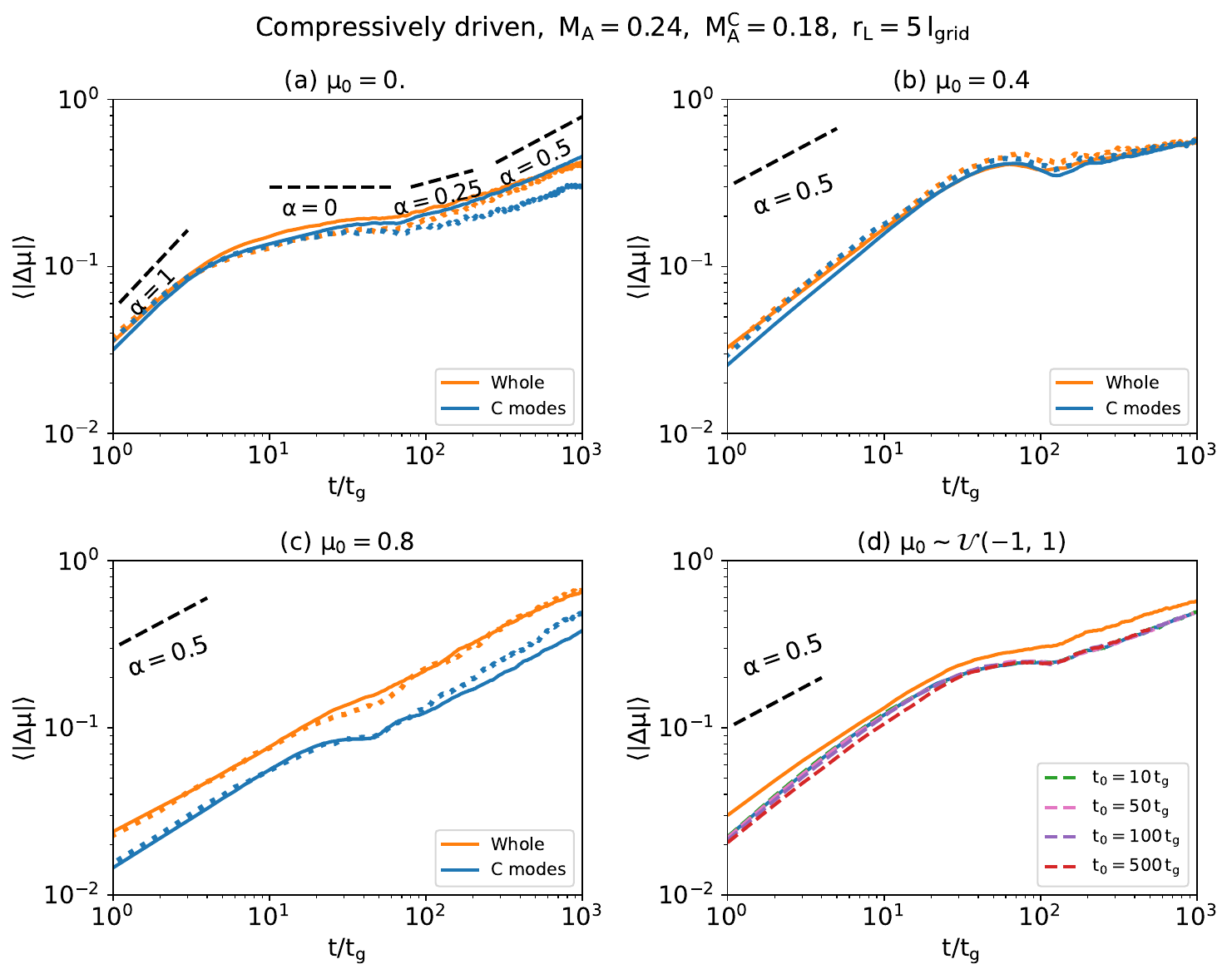}
\centering
\caption{Time evolution of the mean pitch-angle increment $\langle \left| \Delta\mu(t) \right| \rangle$ for particles with different initial $\mu_0$ in compressively driven turbulence of $M_{\rm A}^{\rm C}=0.18, \ M_{\rm A}=0.24$, with $r_{\rm L}=5 \, l_{\rm grid}$. Solid lines show the test-particle results, while dotted lines show the GC Langevin results. (a) Initial pitch angle $\mu_0=0$, (b) $\mu_0=0.4$, (c) $\mu_0=0.8$, (d) $\mu_0$ is sampled from a uniform distibution $\mathcal{U}(-1,1)$. Colored dashed lines showcase different reference start time $t_0$.}
\label{fig:delta_mu}
\end{figure*}

\subsection{Separating mirror interactions from gyroresonance}
\label{subsec:mode_separation}

We characterize pitch-angle transport using the ensemble spread of the pitch-angle cosine, $\langle \left| \Delta\mu(t) \right| \rangle =
\langle \left| \mu(t+t_0)-\mu(t_0) \right| \rangle$, where $t_0$ is the reference start time (by default $t_0=0$; $\mu_0 \equiv \mu(0)$). We quantify its scaling $\langle\Delta\mu\rangle \propto t^{\alpha}$. Here $\alpha=0.5$ corresponds to normal diffusion, $\alpha<0.5$ to subdiffusion, and $\alpha>0.5$ to superdiffusion (with $\alpha=1$ indicating ballistic growth). Time is normalized to the gyroperiod $t_{\rm g}$. To isolate the physical channels that drive $\mu$ evolution, we exploit the orthogonality of Alfv\'enic and compressible perturbations in $\mathbf{k}$-space and decompose the turbulence into Alfv\'enic (A) and compressible (C) components\citep{2002CL, 2003Cho}.

In the C component, mirror interactions and gyroresonance coexist. Mirror is controlled by $\nabla|B|$ and is inherently broadband across the inertial range, whereas gyroresonance is concentrated near cyclotron-resonant scales. Notably, mirror interactions do not rely on a specific resonant scale and exhibit no intrinsic rigidity dependence (Eq.\ref{eq:LGV_mu_sde}), so varying the Larmor radius primarily affects gyroresonant scattering. To construct mirror-dominated reference cases, we suppress gyroresonance by artificially reducing the Lamor radius to a sufficiently small value (e.g. $r_{\rm L} = 1 \, l_{\rm grid}$), thereby shifting the resonant scale 
into the dissipation range. Comparing results from the full turbulence, the mode-decomposed fields and these gyroresonace-suppressed runs allows us to disentangle deterministic mirror confinement from stochastic resonant decorrelation.

In the A component, the perturbations are predominantly transverse to the mean field $\mathbf{B}_0$. As a result, the mirror-relevant gradients $\nabla|B|$ enter only at higher order. Mirror/TTD effects therefore vanish at leading order and the A-modes contribution to pitch-angle scattering is dominated by gyroresonance. However, gyroresonant scattering by Alfv\'enic turbulence is inefficient near $\mu\simeq 0$. Indeed, we see that in large pitch angle cases (small $\mu_0$) the $\mu$ evolution resulted from the C component is nearly identical to that from the full turbulence (Fig.~\ref{fig:delta_mu}a,b).\footnote{We do not present the scattering results for pure Alfv\'enic modes, as the lack of mirror interaction and the Alfv\'enic spectral anisotropy leads to inefficient pitch-angle scattering, rendering the results sensitive to numerical integration accuracy.} This implies that the A component contributes negligibly in this regime. This behavior is further strengthened by the scale-dependent anisotropy of Alfv\'enic turbulence, which suppresses power at the parallel wavenumbers $k_\parallel$ required for the resonance with particles in realistic astrophysical environment\citep{GS95, 2002YL, 2004Yan}.

\begin{figure*}[htbp]
\includegraphics[width=\textwidth]{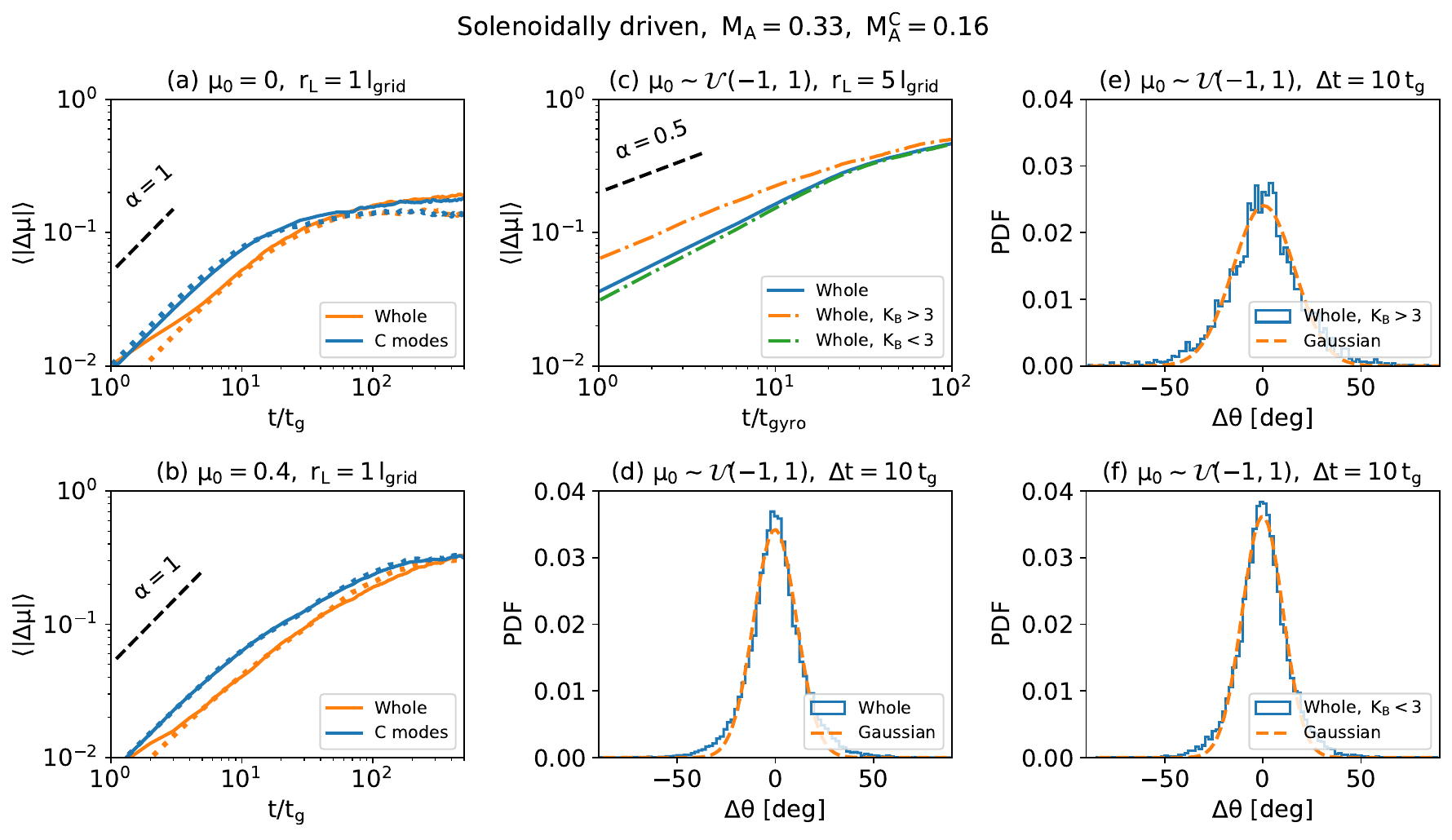}
\centering
\caption{Statistics of particles in solenoidally driven turbulence of $M_{\rm A}=0.33, \ M_{\rm A}^{\rm C}=0.16$. (a) Time evolution of the mean pitch-angle increment $\langle \left| \Delta\mu(t) \right| \rangle$ of particles with $\mu_0=0$ and $r_{\rm L}=1 \, l_{\rm grid}$, so as to suppress gyroresonance and isolate the effects of magnetic mirroring and field-line curvature. Solid lines show the test-particle results, while dotted lines show the GC Langevin results. $\langle \Delta \mu \rangle$ rises faster initially in the C modes than in the whole datacube, because A modes increase $|B|$ without contributing to $\nabla_s B$, thus slightly reducing mirroring $\propto \nabla_s B / |B|$. (b) same as panel (a), but for $\mu_0=0.4$. (c) $\langle \left| \Delta\mu(t) \right| \rangle$ of particles with $\mu_0$ isotropically injected from a uniform distibution $\mathcal{U}(-1,1))$ and $r_{\rm L}=5 \, l_{\rm grid}$. Dash-dotted lines show results of particles injected at high-kurtosis regions and low-kurtosis regions separately. (d) Probability density distributions (PDF) of pitch angle increment $\Delta \theta$ of test-particle simulations for $\mu_0 \sim \mathcal{U}(-1,1)),\ \Delta t = 10 \, t_{\rm g}$. The dash-dotted curve shows the best-fit Gaussian distribution. (e) same as panel (d), but for particles injected at high-kurtosis regions. (f) same as panel (d), but for particles injected at low-kurtosis regions.}
\label{fig:delta_mu_sol}
\end{figure*}

Another question is whether field-line curvature or Richardson superdiffusion of A modes could provide an additional contribution \citep{2014Lazarian, 2022Maiti}. To assess the combined effect of field-line curvature and mirror interactions, we performed a simulation using the solenoidally driven datacube with $r_{\rm L} = 1 \, l_{\rm grid}$. We see the C modes and whole datacube result in nearly indentical $\langle \left| \Delta\mu \right| \rangle$ behaviour (Fig.~\ref{fig:delta_mu_sol}a,b), indicating that A modes field line curvatures are also negligible.

\subsection{Low-$M_{\rm A}$: compound (multi-stage) pitch-angle transport}
\label{subsec:lowMA}

Fig.~\ref{fig:delta_mu} shows $\langle \left| \Delta\mu \right| \rangle$ for a sub-Alfv\'enic compressible run ($M_{\rm A}^{\rm C}=0.18$ for the C components, $M_{\rm A}=0.24$ for the whole datacube) with different initial $\mu_0$. For each $\mu_0$, we inject an ensemble of 5000 particles randomly throughout the turbulence volume. This setup is not intended to model a specific injection scenario. Instead, it serves as a diagnostic of the $\mu$-dependent pitch-angle evolution, noting that particles span a broad range of $\mu$ at any time. A clear regime dependence emerges. Particles injected near $\mu_0=0$ (Fig.~\ref{fig:delta_mu}a) do not enter normal diffusion immediately, but exhibit a four-stage evolution with successive scaling exponents $\alpha\simeq \rm  1, \, 0, \, 0.25$ and 0.5. This sequence reflects the competition between conservative mirror confinement and weak gyroresonant decorrelation. At early times, coherent mirror forcing produces an apparent ballistic growth ($\alpha\simeq 1$), followed by a plateau ($\alpha\simeq 0$) associated with repeated reflections and confinement within a narrow $\mu$ band. At intermediate times, rare resonant kicks gradually accumulate and erode phase coherence of trapped orbits, leading to subdiffusion ($\alpha<0.5$). Only at later time, after particles escape local mirror wells and substantially deviate from their initial $\mu_0$, $\langle\Delta\mu\rangle$ approaches normal diffusion ($\alpha\simeq 0.5$). Mirror interactions therefore play a dual role. They enable particles to cross the $90^{\circ}$ barrier via drift, moving them out of the $90^\circ$ 
inefficient scattering regime and thereby enhancing gyroresonant scattering and escape, while also producing intermittent trapping that delays the onset of the diffusive regime.

This interpretation is supported by the gyroresonance-suppressed runs: the early ballistic and plateau phases closely track the gyroresonance-suppressed case (Appendix Fig.~\ref{fig:delta_mu_rL1grid}), indicating that they are controlled primarily by mirror dynamics, while the subsequent transition from subdiffusion to diffusion requires the cumulative effect of gyroresonant scattering. Thus, in the low-$M_{\rm A}$ regime, escape from mirror potentials naturally produces anomalous scaling over extended times before the diffusive limit is reached.

In contrast, particles injected at small pitch angle, $\mu_0=0.8$ (Fig.~\ref{fig:delta_mu}c) exhibit normal diffusion throughout the entire evolution. Nevertheless, non-Markovianity at large pitch angles can regulate the global transport, because particle pitch angles evolve dynamically and undergo repeated trapping and escape episodes. At $\mu_0=0.4$, the mirror-induced anomalous transport already becomes pronounced as shown in Fig.~\ref{fig:delta_mu}b). Therefore, an anomalous diffusion regime can be also seen in the evolution of $\langle \left| \Delta\mu(t) \right| \rangle$ for an isotropic pitch angle distribution (Fig.~\ref{fig:delta_mu}d). 

\begin{figure*}[htbp]
\includegraphics[width=\textwidth]{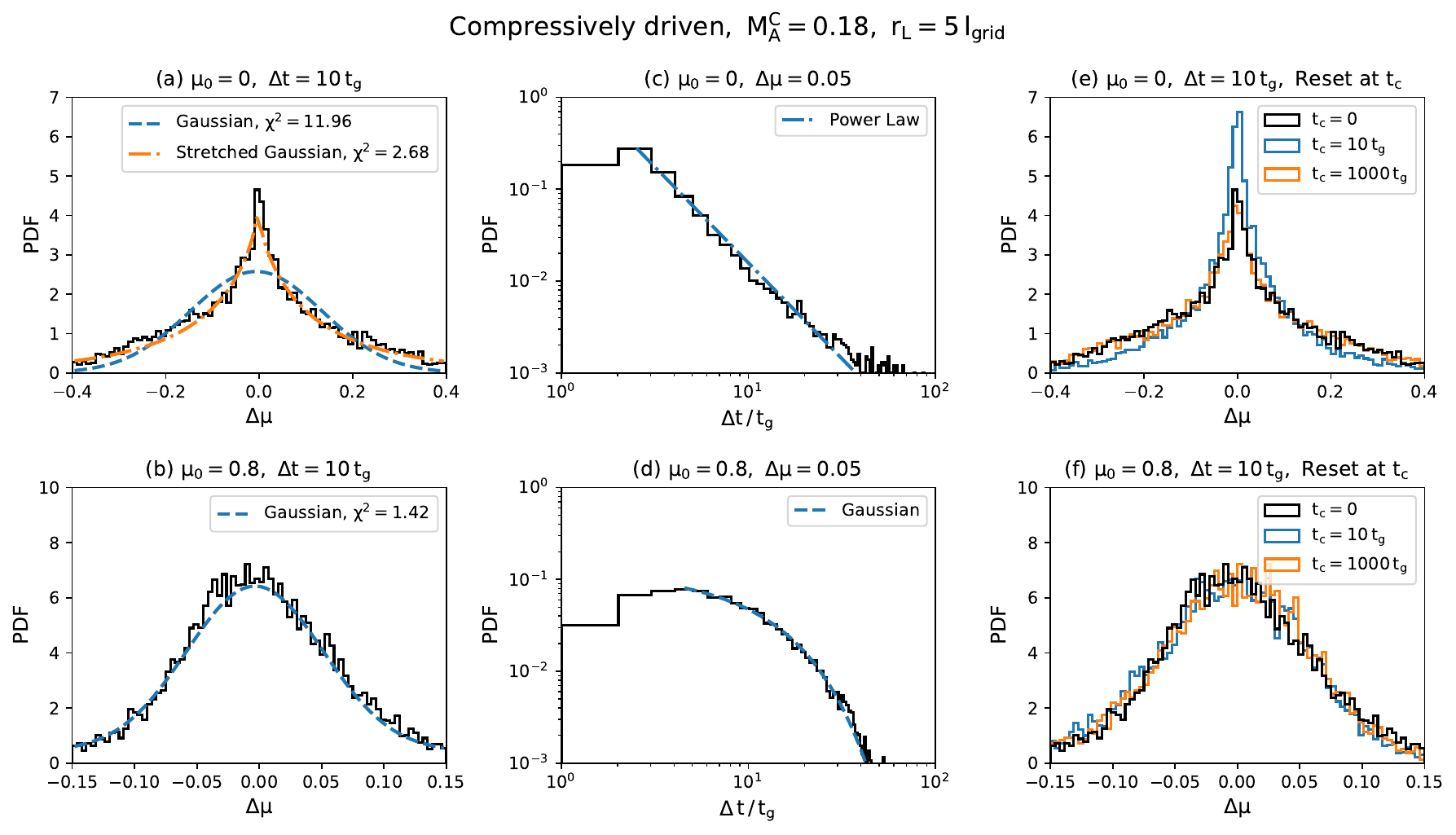}
\caption{Statistical properties of $\Delta\mu$ measured from test-particle simulations in a sub–Alfv\'enic compressive turbulence with $M_{\rm A}^{\rm C}=0.18$. (a) Probability density distributions (PDF) of pitch angle cosine increment $\Delta \mu$ of test-particle simulations for $\mu_0=0, \Delta t = 10 \, t_{\rm g}$. The dashed and dash-dotted curves show fits with a stretched Gaussian and a Gaussian distribution, respectively. The legend lists the reduced chi-square $\chi^2$ for each fit. (b) Same as (a), but for $\mu_0=0.8$. (c) Waiting time distribution for pitch angle cosine increment reaches a threshold $|\Delta \mu| = 0.05$ for $\mu_0 =0$. (d) Same as (b), but for $\mu_0=0.8$. (e) PDF of $\Delta\mu$ for $\mu_0=0$ and $\Delta t=10 \,t_{\rm g}$, computed after resetting $\mu(t_{\rm c})=\mu_0$ at different reference times $t_{\rm c}$, i.e. $\Delta\mu = \mu(t_{\rm c}+\Delta t)-\mu(t_{\rm c}).$ (f) Same as (e), but for $\mu_0 =0.8$.}
\label{fig:PDF}
\end{figure*}

To quantify non-Markovian nature of pitch-angle transport, we further analyze the statistical distribution of the pitch-angle increments $\Delta\mu=\mu(\Delta t)-\mu_0$ with two representative cases, $\mu_0=0$ and $\mu_0=0.8$ (Fig.~\ref{fig:PDF}). For $\mu_0=0$, the increment probability density distributions (PDF) at $\Delta t=10\,t_{\rm g}$ exhibits leptokurtic and is poorly described by a single Gaussian. It is well fit by a stretched Gaussian\citep{2004Kupferman},
\begin{equation}
P(\Delta\mu)=\frac{\kappa}{2\sigma\,\Gamma(1/\kappa)}
\exp\!\left[-\left(\frac{|\Delta\mu|}{\sigma}\right)^{\kappa}\right],
\label{eq:stretched_gaussian}
\end{equation}
where $\sigma$ sets the characteristic width and $\kappa$ controls the tail weight ($\kappa=2$ recovers a Gaussian; $\kappa<2$ yields leptokurtic). For the $\mu_0=0$ case we obtain the best-fit shape parameter $\kappa \simeq 0.72$ (Fig.~\ref{fig:PDF}a), indicating non-Gaussian increments stemming from dynamics associated with mirror trapping and rare escape events. In contrast, for $\mu_0=0.8$ the increment PDF is close to Gaussian (Fig.~\ref{fig:PDF}b), indicating rapid decorrelation and approximately Markovian diffusion.

To further quantify the temporal statistics, we measure the first-passage time $\Delta t$ required for $|\Delta\mu|$ to reach a threshold $|\Delta\mu|=0.05$, which serves as a representative increment rather than a unique choice. The shape of the tail encodes the underlying transport regime. For $\mu_0=0$, the waiting-time distribution exhibits a power-law tail (Fig.~\ref{fig:PDF}c), implying a broad distribution of residence times and a non-Markovian process driven by localized mirror trapping with sporadic gyroresonant escape. For $\mu_0=0.8$, the distribution shows a rapid cutoff (Fig.~\ref{fig:PDF}d), consistent with the Markovian process. 

Memory effects are probed directly by ``reset'' experiments: at a reference time $t_{\rm c}$, we reset $\mu(t)$ to the injected value $\mu_0$ while keeping the spatial position unchanged, and then re-measure the increment statistics $\Delta\mu=\mu(\Delta t + t_{\rm c})-\mu_0$. For $\mu_0=0$, the PDF depends strongly on $t_{\rm c}$ at early times (Fig.~\ref{fig:PDF}e) and relaxes only after long evolution ($t_{\rm c}\sim 10^3\,t_{\rm g}$). This behavior reflects mirror-induced spatial focusing: particles initially distributed randomly are rapidly trapped in localized magnetic wells, and only escape after accumulating sufficient gyroresonant scattering to restore spatial randomness. In contrast, for $\mu_0=0.8$, the PDFs collapse rapidly for different $t_{\rm c}$ (Fig.~\ref{fig:PDF}f), demonstrating fast decorrelation. Together, these diagnostics identify a non-Markovian, anomalous diffusion regime at low $M_{\rm A}$.

\subsection{High-$M_{\rm A}$: rapid decorrelation and diffusive transport}
\label{subsec:highMA}

\begin{figure}[htbp]
\includegraphics[width=0.45\textwidth]{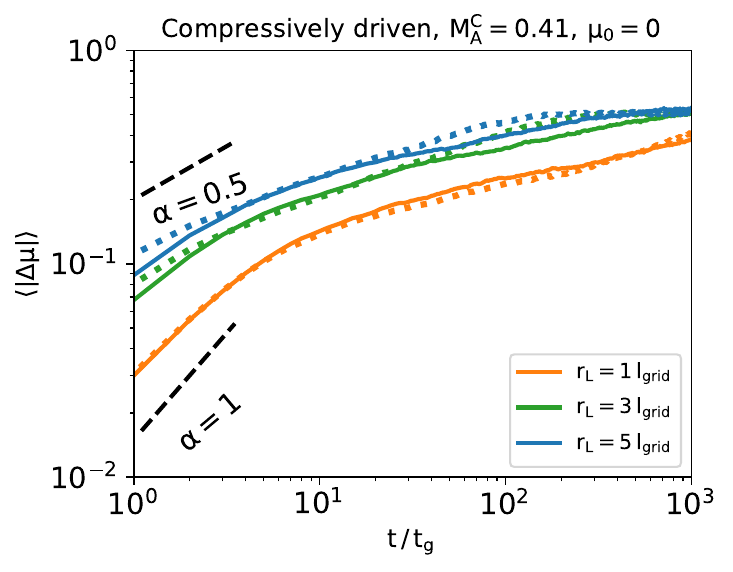}

\caption{Time evolution of the mean pitch-angle increment $\langle \left| \Delta\mu(t) \right| \rangle$ for particles with initial $\mu_0=0$ in compressively driven turbulence of $M_{\rm A}^{\rm C}=0.41$. Solid lines show the test-particle results, while dotted lines show the GC Langevin results. The gyroperiod $t_g$ depends on Larmor radii $R_L$, and thus varies for different particle populations here.}

\label{fig:delta_mu_rL}
\end{figure}

At high $M_{\rm A}$, the pitch-angle dynamics remain dominated by the competition between intermittent mirror trapping and gyroresonant escape. Yet gyroresonant scattering becomes sufficiently efficient that most particles rapidly escape local magnetic potential wells. Within a few gyroperiods, the pitch-angle distribution becomes randomized, and the memory of the injected $\mu_0$ is quickly lost. This holds even for $\mu_0=0$, due to broadened/nonlinear gyroresonance effects \citep{2008Yan} and to the stronger mirror-induced pitch-angle drift, which can sweep particles into regimes where gyroresonant scattering is more effective. While non-Markovianity can still occur for a subpopulation of particles at large pitch angles, rapid escape and efficient decorrelation result in the ensemble-averaged $\langle \Delta\mu\rangle$ dynamics transition to a diffusive scaling $\langle \Delta\mu\rangle\propto t^{\alpha}$ with $\alpha\simeq 0.5$ (red curves in Fig.~\ref{fig:delta_mu_rL}).

To further disentangle gyroresonance from mirror effects, we vary the strength of gyroresonant scattering by changing the particle Larmor radius. Our benchmark choice is $r_{\rm L}=5\,l_{\rm grid}$, for which the resonant wavenumbers lie in the inertial range and gyroresonance is well resolved. We then progressively reduce the Larmor radius to $r_{\rm L}=3$ and $1 \, l_{\rm grid}$, shifting the resonance toward smaller scales and effectively suppressing gyroresonant interactions. As shown in Fig.~\ref{fig:delta_mu}f, when $r_{\rm L}$ is sufficiently small, $\langle \Delta\mu\rangle$ departs from the diffusive law and exhibits anomalous scaling. This controlled simulation demonstrates that the anomalous regime is rooted in prolonged mirror trapping with inefficient resonant escape. When particles escape from local mirror wells quickly, they traverse multiple stochastic mirror structures and the cumulative effect of these encounters enhances scattering and accelerates decorrelation. In this regime, mirror interactions mainly act as an additional broadband scattering enhancement and the transport becomes effectively diffusive.

In the full diffusive limit, we can start from the mirror force update in Eq.~\ref{eq:LGV_mu_sde} and evaluate $D_{\rm \mu\mu}$ via the TGK formula\citep{1957Kubo,2004Shalchi}. If we restrict to GC trajectories and overlook the fast gyromotion, then the orbit phase factor reduces to $\exp(i k_\parallel v \mu t)$.  
In this case there is no dependence on the gyro-orbit. The mirror force only couples to gradients of $B_\parallel$ along the field line, and thus does not require information about the gyrophase. Accordingly, only the $n=0$ resonance survives, and no Bessel functions appear.  
The $D_{\rm \mu\mu}^{\rm Mirror}$ is then calculated as
\begin{equation}
D_{\rm \mu\mu}^{\rm Mirror} =
\frac{v^2 (1 - \mu^2)^2}{4 B_0^2}
\int d^3 k \, k_\parallel^2 \,
P_\parallel^{(M)}(\mathbf{k}) \,
R_0.
\end{equation}
, as detailed in Appendix C. While the standard TTD contribution to pitch-angle diffusion is written as \cite[see][]{1975Volk,2008Yan}
\begin{equation}
D_{\rm \mu\mu}^{\mathrm{TTD}} =
\frac{\Omega^2 (1 - \mu^2)}{B_0^2}
\int d^3 k \,
R_0\,
\frac{k_\parallel^2}{k^2} \,
\left[J_0^{\prime}(W)\right]^2 \,
I^{(M)}(\mathbf{k}),
\label{eq:Duu_mirror}
\end{equation}
In the regime of $W \ll 1$, the Bessel functions can be expanded as $ |J_0^{\prime}(W)| \;\simeq\; \frac{W}{2}\,$
so that the TTD form reduces exactly to the GC mirror result

\begin{equation}
\begin{aligned}
D_{\rm \mu\mu}^{\mathrm{TTD}}
&\xrightarrow{W\ll 1}
\frac{\Omega_0^2 (1 - \mu^2)}{B_0^2}
\int d^3 k \,
R_0\,
\frac{k_\parallel^2}{k^2}\,
\left(\frac{W}{2}\right)^2
I^{(M)}(\mathbf{k})
\\
&=
\frac{v_\perp^2 (1 - \mu^2)}{4 B_0^2}
\int d^3 k \,
k_\parallel^2\,
R_0\,
\frac{k_\perp^2}{k^2}\,
I^{(M)}(\mathbf{k})
\\
&=
\frac{v^2 (1-\mu^2)^2}{4 B_0^2}
\int d^3 k \,
k_\parallel^2\,
R_0\,
P_\parallel^{(M)}(\mathbf{k})
\end{aligned}
\end{equation}
where the $P_\parallel^M$ is the parallel magnetic-power spectrum, and $I^M$ denotes the scalar magnetic-power spectrum. Thus the GC mirror interpretation is independent of the detailed gyro-orbit, sensitive to all compressive $\delta B_\parallel$ fluctuations, and equivalent to TTD in the small $k_{\perp}$ limit, which corresponds to either long wavelength or quasi-slab modes. The latter is expected for fast modes in low-$\beta$ plasmas due to collisionless damping \citep{2004Yan, 2008Yan, 2006Petrosian, Suzuki2006, 2024Zhao, 2025Hou, 2025Zhao}.

\subsection{Mean free path from $D_{\rm \mu \mu}$ and $D_{\parallel}$}

The parallel mean free path \( \lambda_\parallel \) can be estimated from pitch-angle diffusion under the assumption of an isotropic pitch-angle distribution, using the following expression\citep{1974Earl, 2022Maiti}:

\begin{equation}
\lambda_\parallel
= \frac{3v}{4}\int_0^1 \frac{(1-\mu^2)^2}{D_{\mu\mu}}\, d\mu,
\qquad
D_{\mu\mu} = \frac{\langle (\mu-\mu_0)^2\rangle}{2t}.
\label{eq:mfp_Duu}
\end{equation}
Note that $D_{\mu\mu}$ here denotes an effective pitch-angle diffusion coefficient that incorporates the combined effects of both gyroresonance and mirror interactions.

Alternatively, the mean free path can also be estimated from the spatial diffusion coefficient\citep{1974Earl, 2022Maiti}:

\begin{equation}
\lambda_\parallel = \frac{3D_\parallel}{c},
\qquad
D_\parallel = \frac{\langle (z-z_0)^2 \rangle}{2t}.
\label{eq:mfp_Dpara}
\end{equation}

These two methods are expected to agree when pitch angle scattering is efficient over the entire $\mu$ domain, so that the distribution of pitch angle $f(\mu)$ relaxes to isotropic. 

In our simulations, this equivalence breaks down whenever scattering is incomplete, such as in the pure Alfv\'enic modes or in the low-$M_{\rm A}$ datacube cases. In these regimes, particles with large pitch angles cannot be sufficiently scattered and do not reach the normal diffusive regime. In this case, a stationary $D_{\rm \mu\mu}$ and $\lambda_\parallel$ cannot be obtained reliably. 

When scattering is sufficient, most notably in the high-$M_{\rm A}$ datacubes, the two estimates of mean free path converge. In this regime, particles quickly drift and escape individual mirror potential wells and decorrelate from initial state. Pitch-angle scattering behavior transitions to a normal diffusion regime within a short time. Mirror/TTD interactions then contribute efficiently to a well-defined $D_{\rm \mu\mu}$, and Eq.~\ref{eq:mfp_Duu} and Eq.~\ref{eq:mfp_Dpara} yield consistent values of the parallel mean free path (Table \ref{tab:mfp_modes}). 

\begin{table}[t]
\centering
\caption{Mean free path $\lambda_{\parallel}$ inferred from two diagnostics of $D_{\mu\mu}$ and $D_{\parallel}$.}
\begin{tabular}{lccc}
\hline\hline
& $M_{\mathrm{A}}=0.51$ & $M_{\mathrm{A}}=0.58$ & $M_{\mathrm{A}}=0.66$ \\
\hline
$\lambda_{\parallel}\,[L_{\rm box}]$ from $D_{\parallel}$
& 1.09 & 0.69 & 0.50 \\
$\lambda_{\parallel}\,[L_{\rm box}]$ from $D_{\mu\mu}$
& 0.96 & 0.80 & 0.55 \\
\hline
\end{tabular}
\label{tab:mfp_modes}
\end{table}

\subsection{Large angle scattering from intermittency}
Some studies have suggested that scattering may be governed by high-order intermittency, which produce rare but large-angle scattering events \citep[e.g.,][]{2024Butsky}. To assess the role of higher-order intermittency, we disentangle contributions from coherent structures and the background cascade using second- and fourth-order structure functions of magnetic increments. Following the procedure in \citet{2025Hou}, we quantify magnetic intermittency through the kurtosis
\begin{equation}
K_{\rm B}(\Delta\mathbf{r}) \;\equiv\; 
\frac{\big\langle \big[\Delta B\big]^4 \big\rangle}{\big\langle \big[\Delta B\big]^2 \big\rangle^{2}},
\label{eq:KB_def}
\end{equation}
where $\Delta B \;\equiv\; B(\mathbf r+\Delta\mathbf r)-B(\mathbf r)$ and $\Delta\mathbf r$ is the spatial separation vector. Values $K_{\rm B}>3$ indicate non-Gaussian, heavy-tailed statistics typical of intermittent structures. We inject particle ensembles into high-$K_{\rm B}$ and low-$K_{\rm B}$ regions separately with isotropically distributed pitch angles and measure the $\langle \left| \Delta\mu \right| \rangle$ behavior. As shown in Fig.~\ref{fig:delta_mu_sol}c-f, particles initialized in high-$K_{\rm B}$ regions exhibit larger $\langle \left| \Delta\mu \right| \rangle$, but these regions occupy a small filling factor (regions with $K_{\rm B} > 3$  occupy about 12.93\% of the total simulation volume), and the overall $\langle \left| \Delta\mu \right| \rangle$ is not dominated by intermittent hotspots, indicating that scattering is governed by broadly distributed, lower-order fluctuations rather than rare intermittency.

\section{Discussion}

\subsection{Negligence of damping effects}
In MHD turbulence generated by Athena++, artificial isotropic viscosity is added at small-scale dissipation ranges to maintain numerical stability and to prevent energy pile up at large $k$ \citep{2020Stone}. In real plasma environment, multiple physical damping channels operate and their relative importance depends on both wavelength scale and plasma parameters. Because interstellar phases differ in density, ionization fraction, and plasma $\beta$, they have different dominant damping mechanisms \citep{2004Yan,YLD2004}. Low density rarefied media such as the solar wind and Galactic halo tend to be governed by collisionless damping. In warm ionized or neutral media, collisional viscosity and thermal conduction gain importance. In partially ionized phases, ion–neutral friction preferentially damps currents transverse to $\mathbf{B}$. In all cases, damping preferentially suppresses large-$k$ power. Moreover, for compressive (mostly fast-mode) fluctuations in low-$\beta$ plasmas, both theory and recent simulations and observations show stronger damping at oblique/quasi-perpendicular angles than in quasi-slab geometries \citep{2006Petrosian, 2008Yan, 2024Zhao, 2025Hou, 2025Zhao}. Including physical damping therefore further suppresses the high-$k$ portion of the spectrum and quasi-perpendicular components, and gradually restores the equivalence between TTD and mirror interactions in the diffusive limit.

\subsection{Anisotropic particle distribution and gyroresonance instability}

In this paper, the possible kinetic feedback is not included. In the case of low $M_A$ turbulence, for instance, the insufficient scattering is bound to drive anisotropy in the particle distribution. The pressure anisotropy can excite gyroresonance instability, which can increase the scattering rate, in turn \citep{Yan2011, lebiga2018}. As the result, particle transport can restore normal diffusion through the self-excited Alfven waves \citep[see][]{Kulsrudbook, Yan2012,Bykov2013}. The interplay between mirror interaction and the gyroresonance counting for the feedback will be studied elsewhere.  

\section{Summary}
In this work, we showed that mirror structures in turbulent magnetic fields drive non-Markovian pitch-angle diffusion at large pitch angles, which influences the transport picture of CRs. By combining full‑orbit test‑particle simulations in 3D MHD turbulence with a guiding-center (GC) Langevin model, we disentangle the roles of gyroresonant scattering and mirror interactions. We find that mirror interactions act multiscale: they are local in real space, generate non-Markovian memory through trapping in mirror potential wells, and statistically produce anomalous pitch angle transport. At large Alfv\'enic Mach number $M_A$, gyroresonant kicks become strong enough to let particles hop across many localized mirror structures, causing decoherence and restoring to normal diffusion. The coupling between these two processes governs the transport of particles. Our main results can be summarized as follows:
\begin{enumerate}
\item Mirror structures, generated primarily by compressive modes (and by pseudo‑Alfv{\'e}n fluctuations in the incompressible limit), introduce a deterministic drift in $\mu$ which naturally breaks the QLT prediction of pitch-angle scattering and bring the particles out of the $90^\circ$ barrier.
\item Gyroresonant scattering modulates the escape from mirror wells and is essential for establishing normal diffusion. The higher the Alfv{\'e}n Mach number $M_{\rm A}$ is, the faster normal diffusion restores.
\item The mirror effect leads to non-Markovian, anomalous pitch-angle diffusion with memory, characterized by platokurtic distributions of pitch angle increment and heavy‑tailed PDF of waiting time.
\item The pitch angle transport is dominated by the interaction with compressible fluctuations. The mean free path calculated with $D_{\mu\mu}$ contributed from the compressible modes agrees well with the measurement from spatial displacement. The contributions from Alfv\'en modes and large angle scattering from the intermittent structures are negligible. 
\item In the diffusive limit, transit time damping (TTD) treatment is adequate for capturing the magnetic mirror interaction in the case of $k$, or $k_\perp \ll 1/r_L$. 
\end{enumerate}

Our results suggest that CR propagation experiences intermittent spatial trapping of particles in magnetic wells, which reacurs whenever the pitch angle becomes sufficient large, and hence is inherently non-Markovian. This effect is more prominent for lower $M_{\rm A}$ case. Such a non-Markovian memory framework is essential for accurately modeling CR confinement, escape, and nonthermal emission in/around CR sources. In realistic astrophysical environment, such as the multiphase interstellar medium, $M_{\rm A}$ may vary over a wide range. Consequently, the balance between compressive mirror trapping and gyroresonant scattering is environment-dependent, leading to distinct CR transport phenomenology across different regions. 


\section{Acknowledgments}
K.Y. acknowledges the National Science Foundation of China under grants No. 125B2062. R.-Y.L. acknowledges the National Science Foundation of China under grants No. 12393852. We acknowledge the computing resources from the high-performance computers at the NHR center NHR@ZIB, with the project No. bbp00080. We acknowledge the use of Chatgpt for assistance with coding and analysis in exploratory discussions, as well as for improving the grammar and clarity of the manuscript.

\appendix

\section{Equivalence of Langevin Description and Focused Fokker-Planck Equation}

From the Langevin model described Eq.~\ref{eq:LGV_mu_sde}, 

\[
\delta \mu=
-\frac{1-\mu^2}{2B}\,v\,\nabla_s B\, \delta t
+\frac{\partial D_{\mu\mu}}{\partial\mu}\,\delta t
+\sqrt{2D_{\mu\mu}}\,\eta_t.
\]
Denote $A = -\frac{1-\mu^2}{2B}\,v\,\nabla_s B
+\frac{\partial D_{\mu\mu}}{\partial\mu}$
where $s$ denotes the direction along the local magnetic field, and taking the  ensemble average, 
$\langle \delta\mu \rangle = A\,\delta t, \
\langle (\delta\mu)^2 \rangle = 2D_{\mu\mu}\,\delta t$.
The probability transfer of the distribution function over a small time step gives

\[
f(\mu,s,t+\delta t)
=
\left\langle
f(\mu-\delta\mu,\; s-\delta s,\; t)
\right\rangle, 
\]
Expand $f(\mu,s,t)$ about $(\mu,s)$ up to second order in $\delta\mu$ and first order in $\delta s$,
\[
f(\mu-\delta\mu, s-\delta s,\; t)=
f(\mu,s,t)
-\delta\mu\,\partial_\mu f(\mu,s,t)
-\delta s\,\partial_s f(\mu,s,t) \\
+\frac{1}{2}(\delta\mu)^2\,\partial_\mu^2 f(\mu,s,t)
+O(\delta t^{3/2}),
\]
which leads to
\[
f(\mu,s,t+\delta t)-f(\mu,s,t)
= -\,\partial_\mu(Af)\,\delta t
+ \partial_\mu^2(D_{\mu\mu}f)\,\delta t -\, v\mu\,\partial_s f\,\delta t .
\]
Dividing by $\delta t$, we obtain
\[
\frac{\partial f}{\partial t}
+ v\mu\,\frac{\partial f}{\partial s}
= -\,\frac{\partial}{\partial \mu}(A f)
+ \frac{\partial^2}{\partial \mu^2}(D_{\mu\mu} f)
=\frac{\partial}{\partial \mu}
\left[
v (1-\mu^{2})\frac{\nabla_s B}{2 B}f
 + D_{\mu\mu}(\mu)\,\frac{\partial f}{\partial \mu} \right]
\]

This conserves the number of particles by integrating over all $s$ and $\mu$. By introducing the substitution $f = g/B$, one obtains
\[
\frac{1}{B}\frac{\partial g}{\partial t}
+v\mu\left(\frac{1}{B}\frac{\partial g}{ \partial s}-\frac{\nabla_s B}{B}\frac{g}{B}
\right)=
\frac{\partial}{\partial\mu}
\left[
v(1-\mu^2)\frac{\nabla_s B}{2B}\frac{g}{B}
+
D_{\mu\mu}\frac{1}{B}\frac{\partial g}{\partial\mu}
\right].
\]
which reduces to the focused Fokker-Planck equation
\[
\frac{\partial g}{\partial t}
+ v\mu\,\frac{\partial g}{\partial s}-
v (1-\mu^{2})\frac{\nabla_s B}{2 B}\frac{\partial g}{\partial \mu}
= \frac{\partial}{\partial \mu}\left( D_{\mu\mu}\,\frac{\partial g}{\partial \mu} \right).
\]

\begin{figure*}[htbp]
\includegraphics[width=0.95\textwidth]{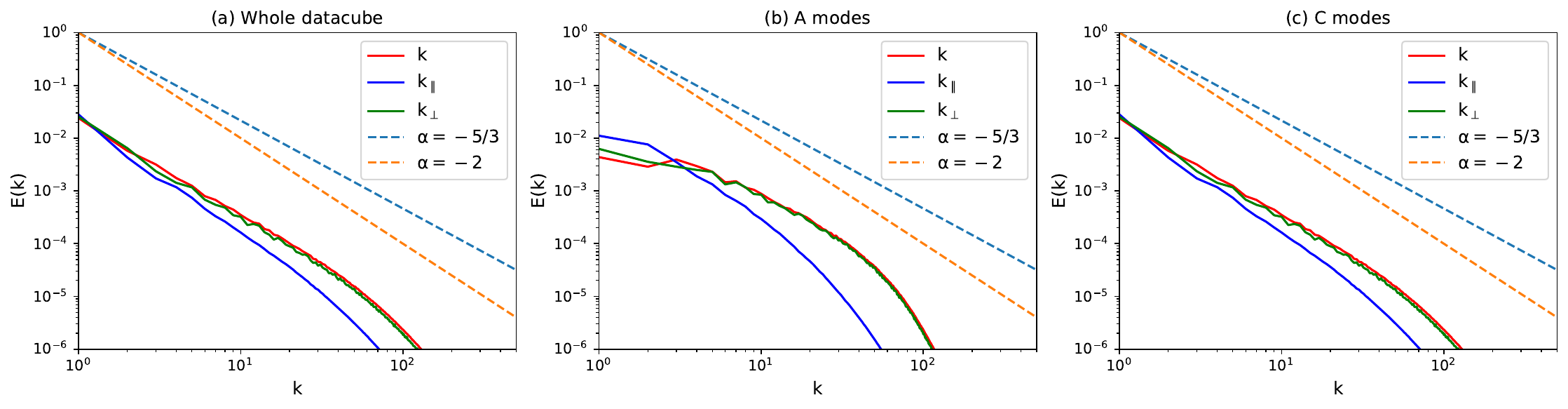}

\caption{Power spectrum density (PSD) for the (a) whole datacube (b) decomposed Alfv\'enic modes (c) decomposed compressible modes.}
\label{fig:PSD}
\end{figure*}

\section{Numerical comparisons of Langevin Description and Focused Fokker-Planck Equation in 1D case}

\begin{figure*}[htbp]
\includegraphics[width=0.45\textwidth]{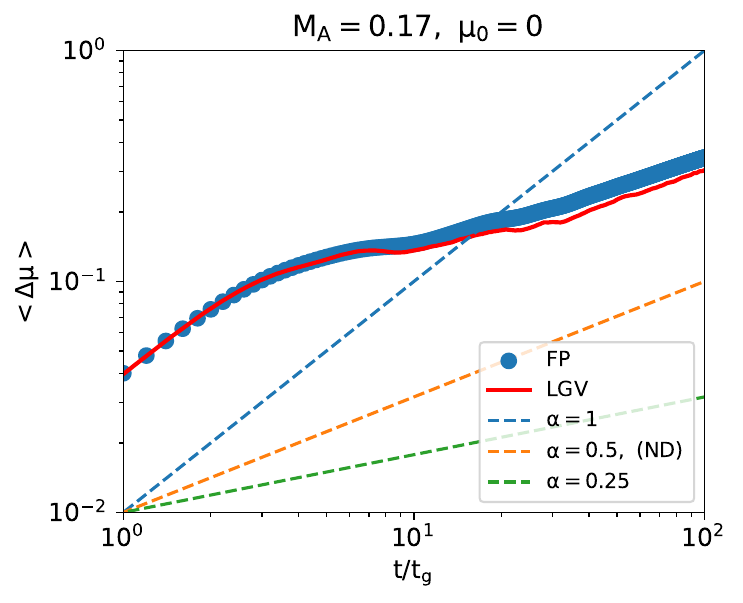}
\includegraphics[width=0.45\textwidth]{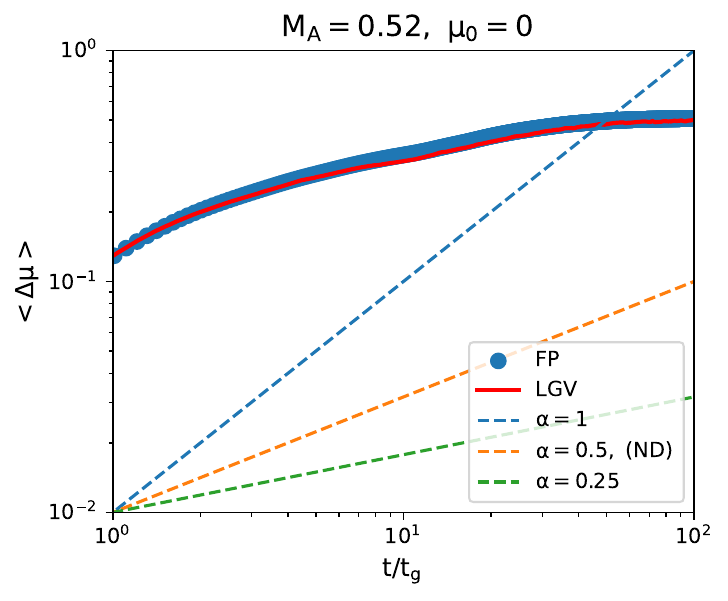}
\caption{One dimensional numerical comparisons of focused Fokker-Planck (FP) solutions and Langevin solutions (LGV) under low-$M_{\rm A}$ case (left panel) and high-$M_{\rm A}$ case (right panel). $\alpha=0.5$ represents the normal diffusion (ND) case.} 
\label{fig:comp_FP_LGV}
\end{figure*}

We present a one-dimensional numerical comparison of Langevin description and the focused Fokker-Planck equation in Fig.~\ref{fig:comp_FP_LGV}. Since solving the FP equation in full 3D MHD turbulence is computationally expensive, we adopt a simplified 1D setup defined along the magnetic field where both approaches can be computed efficiently and compared directly. We construct a set of synthetic, gradient-varied magnetic fields $B(z)$ by superposing Fourier modes with random phases and a spectrum $E(k)\propto k^{-2}\exp(-k/50),$
and evaluate the corresponding focusing term. Here, the $k^{-2}$ dependence is adopted to represent a spectrum measured along the local magnetic-field direction\citep{2015Beresnyak}, while the exponential factor is included as an effective high-$k$ dissipation cutoff. To approximate gyroresonant scattering, we set input diffusion coefficients
$D_{\mu\mu}^{\rm G} = D_0(1-\mu^2)|\mu|.$
We then solve the focused FP equation numerically and the Langevin equation with the same $B(z)$ and $D_{\mu\mu}^{\rm G}$. Two representative parameter sets are considered: $M_{\rm A}=0.17$ with $D_0=0.005\,\Omega_0$, and $M_{\rm A}=0.52$ with $D_0=0.04\,\Omega_0$, where the gyrofrequency $\Omega_0$ is used to normalize the scattering rate. In both cases, the Langevin results are consistent with the FP solution, confirming the numerical equivalence of the two descriptions.

\begin{figure*}[htbp]
\includegraphics[width=\textwidth]{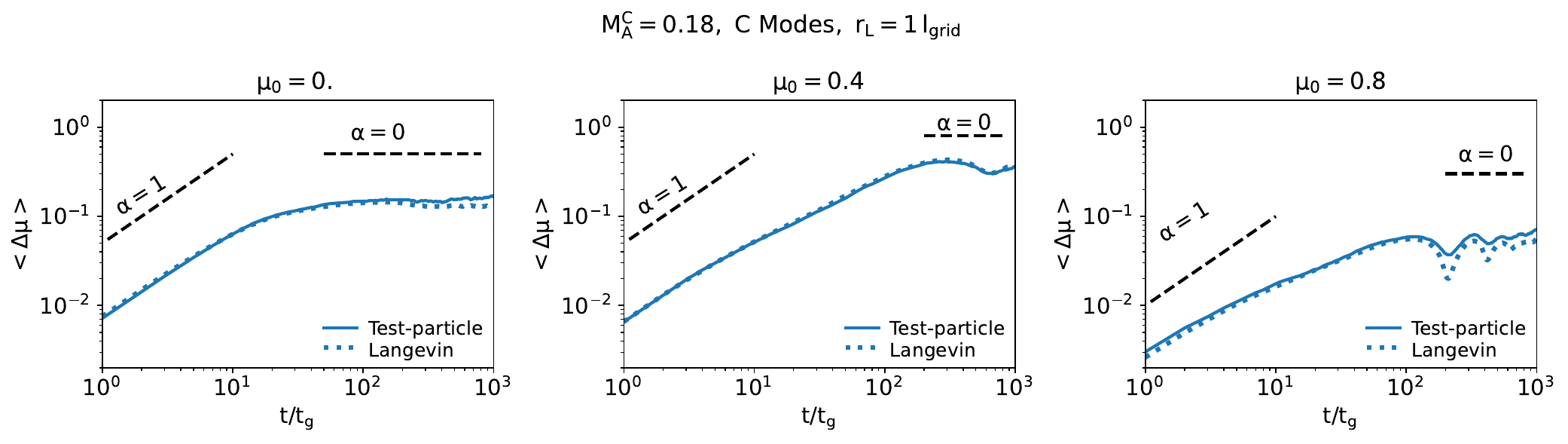}

\caption{Time evolution of the mean pitch–angle increment $\langle\Delta\mu\rangle$ for particles with different initial $\mu_0$ in a sub–Alfv\'enic compressible run ($M_{\rm A}^{\rm C}=0.18$) with $r_L=1 \, l_{\rm grid}$. With an extremely small gyroradius, Gyroresonance is suppressed in this case.}
\label{fig:delta_mu_rL1grid}
\end{figure*}

\section{Markovian limit of the mirror interactions}

If magnetic compressions experienced by the particles decorrelate sufficiently rapidly, mirror interactions can be coarse-grained into an effective stochastic process. In that limit, the relevant quantity is the pitch-angle diffusion coefficient $D_{\mu\mu}$ where \citep[see][]{1957Kubo,2004Shalchi}
\[
D_{\mu\mu}^{\rm Mirror} = \int_0^\infty d\tau
\left\langle \dot{\mu}_{\rm M}(t)\dot{\mu}_{\rm M}^*(t+\tau)
\right\rangle .
\]
Using
\[
\dot{\mu}_{\rm M} =  -\frac{v(1-\mu^2)}{2B_0}\nabla_{\parallel} \delta B_{\parallel},
\]
one obtains

\[
D_{\mu\mu}^{\rm Mirror} = 
\frac{v^2(1-\mu^2)^2}{4B_0^2}
\int_0^\infty 
\left\langle
\nabla_{\parallel}\delta B_{\parallel}(t)
\nabla_{\parallel}\delta B_{\parallel}^*(t+\tau)
\right\rangle d\tau,
\]
after Fourier transform $
\delta B_{\parallel}(\mathbf{x}, t) = \int 
\delta B_{\parallel}(\mathbf{k}, t) e^{i\mathbf{k}\cdot\mathbf{x}} d^{3}k , \
\nabla_{\parallel}\delta B_{\parallel} \rightarrow i k_{\parallel} \delta B_{\parallel}(\mathbf{k})
$, 
\begin{equation}
\begin{aligned}
\left\langle
\nabla_{\parallel}\delta B_{\parallel}(t)
\nabla_{\parallel}\delta B_{\parallel}^*(t+\tau)
\right\rangle
&= \int d^3k \, d^3k' \,
k_{\parallel} k'_{\parallel}
\left\langle
\delta B_{\parallel}(\mathbf{k}, t)
\delta B_{\parallel}^*(\mathbf{k}', t+\tau)
e^{i(\mathbf{k}-\mathbf{k}')\cdot \mathbf{x}}
\right\rangle \\
&\simeq
\int d^3k \, d^3k' \, k_{\parallel} k'_{\parallel}
\left\langle \delta B_{\parallel}(\mathbf{k}, t)
\delta B_{\parallel}^*(\mathbf{k}', t+\tau)\right\rangle
\left\langle e^{i(\mathbf{k}-\mathbf{k}')\cdot \mathbf{x}}\right\rangle \\
&=
\int d^3k \, d^3k' \, k_{\parallel} k'_{\parallel}
P_{\parallel}(k) \, \delta(\mathbf{k}-\mathbf{k}')
\left\langle e^{i(\mathbf{k}-\mathbf{k}')\cdot \mathbf{x}} \right\rangle \\
&= \int d^3k \, k_{\parallel}^2 \, P_{\parallel}(k)
\left\langle e^{i\mathbf{k} \mathbf{x}} \right\rangle .
\end{aligned}
\end{equation}
\citep[see][]{2004Shalchi}. This leads to
\[
D^{\mathrm{Mirror}}_{\mu\mu} =
\frac{v^{2}(1-\mu^{2})^{2}}{4 B_{0}^{2}}
\int d^{3}k \, k_{\parallel}^{2} P_{\parallel}(\mathbf{k})
\int_{0}^{\infty} \left\langle e^{i\mathbf{k} \mathbf{x}} \right\rangle d\tau 
\]

Consider the guiding center has a Gaussian distribution along the field line following \citet{2008Yan},
\[
\left\langle e^{i\mathbf{k} \mathbf{x}} \right\rangle = e^{ik_{\parallel}\left\langle z \right\rangle} e^{-k_{\parallel}^2 \sigma_z^2/2}
\]

\[
\sigma_z^2
=
\left\langle \Delta v_{\parallel}^2 \right\rangle t^2
=
v_{\perp}^2
\left(
\frac{\langle \delta B_{\parallel}^2 \rangle}{B_0^2}
\right)^{1/2}
t^2.
\]

This gives the zeroth-order nonlinear broadened resonance function

\begin{equation}
\begin{aligned}
R_0
&= \int_{0}^{\infty} \left\langle e^{i\mathbf{k} \mathbf{x}} \right\rangle d\tau \\
&= \Re \int_{0}^{\infty} dt \,
\exp\left[
ik_{\parallel} v_{\parallel} t
-\frac{1}{2} k_{\parallel}^2 v_{\perp}^2 t^2
\left(
\frac{\langle \delta B_{\parallel}^2 \rangle}{B_0^2}
\right)^{1/2}
\right] \\
&=
\frac{\sqrt{\pi}}{|k_{\parallel}\Delta v_{\parallel}|}
\exp\left[
-\frac{\left(k_{\parallel} v\mu \right)^2}
{k_{\parallel}^2 \Delta v_{\parallel}^2}
\right] \\
&\simeq
\frac{\sqrt{\pi}}{|k_{\parallel}| v_{\perp} M_A^{1/2}}
\exp\left[
-\frac{\mu ^2}
{(1-\mu^2) M_A}
\right].
\end{aligned}
\end{equation}

Thus we obtain the expression of effective mirror-induced diffusion coefficient in Eq.~\ref{eq:Duu_mirror}
\[
D^{\mathrm{Mirror}}_{\mu\mu} =
\frac{v^{2}(1-\mu^{2})^{2}}{4 B_{0}^{2}}
\int d^{3}k \, k_{\parallel}^{2} P_{\parallel}(\mathbf{k})
R_0.
\]




\end{document}